# Preliminary study on the impact of EEG density on TMS-EEG classification in Alzheimer's disease

Alexandra-Maria Tăuțan[1,2,*], Elias Casula[3], Ilaria Borghi[3], Michele Maiella[3], Sonia Bonnì[3], Marilena Minei[3], Martina Assogna[3], Bogdan Ionescu[1], Giacomo Koch[3,4] and Emiliano Santarnecchi[2]

*Abstract*— Transcranial magnetic stimulation co-registered with electroencephalographic (TMS-EEG) has previously proven a helpful tool in the study of Alzheimer's disease (AD). In this work, we investigate the use of TMS-evoked EEG responses to classify AD patients from healthy controls (HC). By using a dataset containing 17AD and 17HC, we extract various time domain features from individual TMS responses and average them over a low, medium and high density EEG electrode set. Within a leave-one-subject-out validation scenario, the best classification performance for AD vs. HC was obtained using a high-density electrode with a Random Forest classifier. The accuracy, sensitivity and specificity were of 92.7%, 96.58% and 88.82% respectively.

*Clinical relevance*— TMS-EEG responses were successfully used to identify Alzheimer's disease patients from healthy controls.

## I. INTRODUCTION

Alzheimer's disease is a neurodegenerative disorder which causes severe cognitive impairments. Its worldwide prevalence is high and is expected to increase in the next decades [1]. As no effective treatment is available, tools developed for its treatment and diagnosis are still needed.

Transcranial Magnetic Stimulation (TMS) used in combination with electroencephalography (EEG) has proven to be a powerful tool in the study of the brain [2]. Furthermore, it has found various applications in the treatment and study of AD pathology [3]. The use of TMS-EEG is particularly interesting as it reveals issues with neuronal communication and transmission of information and Alzheimer's disease is often characterized as a disconnection syndrome.

Although TMS has previously been used to characterize AD, few works investigated its diagnostic and differential diagnostic potential. For instance, Benussi et al. [4] uses motor responses obtained after applying TMS to classify AD patients from dementia with Lewy bodies and Frontotemporal dementia. Using a Radom Forest classifier, classification accuracies ranging from 0.89 to 0.92 were obtained. No works have been identified that use TMS in combination with recorded EEG response to classify AD patients.

In this study, we aim to: (i) use of TMS-evoked EEG potentials (TEPs) to classify AD patients from healthy controls (HC), (ii) investigate the use of a low, medium or high electrode set as input, (iii) analyze the impact of using various classification algorithms.

This paper is organized as follows. Section II provides an overview of the methods used in classification of AD patients and the experiments performed. Section III details the results, while Section IV provides a brief analysis of the results. Section V concludes the paper.

## II. METHODS

Our approach for classifying Alzheimer's disease patients from HC using TMS-EEG data is outlined in the six steps from the diagram presented in Figure 1. The TMS-EEG dataset is manually pre-processed for the removal of artefacts. Features are extracted per trial and averaged globally based on different electrode montage configurations. The averaged values are used as input to a classification algorithm which creates a model for distinguishing the two classes. In the final step, the result is evaluated. The following sections detail the six steps.

### A. Dataset

A total of 17 AD and 17 HC TMS-EEG recordings were included in the experiments (Fig. 1, step 1). The participants were aged matched with a mean age of 72.35±7.72 years and 59% females for the Alzheimer's group and 71.11±6.28 years and 59% females in the healthy control group. All subjects were evaluated for AD using clinical biomarkers and neuropsychological tests for cognitive assessment prior to enrollment according to the latest AD diagnostic criteria [5]. The average mini mental state examination result was 17.29±6.23 for AD patients and 29.06±1.34 for HC.

The data collection took place at the Santa Lucia Foundation in Rome, Italy. The protocol for data collection was approved by the ethics committee of the Santa Lucia Foundation and was conducted according to the principles outlined in the Declaration of Helsinki. A written consent was signed either by the participant or their legal representative.

EEG data was collected from 62 TMS-compatible Ag/AgCl electrodes positioned according to the 10-10 system nomenclature locations. Data was collected with a sampling rate of 5kHz, while ensuring that the skin-electrode impedance was below 5 $k\Omega$ for the duration of the recording.

[1]Alexandra-Maria Tăuțan and Bogdan Ionescu are with University Politehnica of Bucharest, AI Multimedia Lab, Research Center CAMPUS, 6 Iuliu Maniu Bd., 061344, Bucharest, Romania (e-mail:bogdan.ionescu@upb.ro)

[2]Alexandra-Maria Tăuțan and Emiliano Santarnecchi are with Precision Neuromodulation Program Network Control Laboratory, Gordon Center for Medical imaging, Department of Radiology, Massachusetts General Hospital, Harvard Medical School, Boston, MA, USA (e-mail:atautan@mgh.harvard.edu, esantarnecchi@mgh.harvard.edu)

[3]Ilaria Borghi, Michele Maiella, Sonia Bonnì, Marilena Minei, Martina Assogna, Elias Casula and Giacomo Koch are with the Santa Lucia Foundation, Via Ardeatina 306-354, 00179, Rome, Italy (e-mail:e.casula@hsantalucia.it, g.koch@hsantalucia.it)

[4]Giacomo Koch is with Department of Neuroscience and Rehabilitation, Section of Human Physiology, University of Ferrara, Via Fossato di Mortara 17-19, 44121 Ferrara, Italy

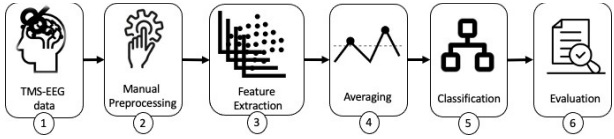

Fig. 1. Steps in the classification of AD and HC from TMS-EEG data.

TMS pulses were applied on the left dorso-lateral prefrontal cortex using a figure of eight coil with a 2.2T output. The localization of the stimulation site was performed based on MRI anatomical references using a pre-established procedure [2]. The stimulation intensity was adjusted based on the individual motor threshold and the distance from the scalp [6]. Pulses were delivered at an inter-stimulus interval of 2-4 seconds in blocks of 120 single-pulses.

*B. EEG Preprocessing*

Due to the nature of the non-invasive scalp EEG signals, multiple sources of artefacts can interfere with the recordings. To analyze the collected data, the artefacts are removed through a manual pre-processing step to avoid confounding factors in the analysis (Fig. 1, step 2) [2]. The recordings were segmented 500ms prior to 1000ms after the TMS pulse. The recording during the application of the pulse was removed and a cubic interpolation was performed. The data was filtered between 1 and 80Hz using a zero-phase Butterworth filter. Afterwards, the data was down sampled to 1kHz. Artefacts such as blinks, eye movements, muscle activity, interfering electrocardiogram and evoked auditory potentials were removed by manually selecting components from several subsequent independent component analysis applied to the data. The manual selection was performed based on criteria already available in literature [7]. Finally, electrodes were referenced to the electrode average.

*C. Feature Extraction*

A total of fourteen features were extracted after EEG signal normalization for each trial after a TMS pulse. An overview of the features is available in Table I.

Metrics based on descriptive statistics can efficiently characterize the EEG signal [8]. Here we use simple metrics such as maximum, minimum and mean amplitude of the EEG segment as well as the third (skewness) and fourth (kurtosis) statistical moment of the distribution of samples.

Hjorth activity, mobility and complexity are time domain based parameters that take into account the variance of the EEG signal and can provide a parallel to frequency characterization [9]. Their definition is provided in Table I.

The energy content of the signal in time domain can also be useful in characterizing the TMS-evoked EEG response.

TEPs have been previously characterized and linked to AD and various other dementias [3]. Several studies have investigated different latencies and peak amplitudes, but a consensus on the proper time windows for analysis has not been reached as there is a significant inter-individual variability in responses to TMS [10]. Based on prior literature, we have selected four windows for peak detection as outlined in Table I. The peak amplitudes are used as input for the classification. An example of the peaks detected in the four windows (P1, P2, P3, P4) is presented in Figure 2. The peaks are polarity independent.

TABLE I
OVERVIEW OF THE FEATURES EXTRACTED FROM THE TEPS.

| | Name | Description |
|---|---|---|
| Descriptive Statistics | Max | Maximum value of the EEG signal amplitude after applying the TMS pulse |
| | Min | Minimum value of the EEG signal amplitude after applying the TMS pulse |
| | Mean | Mean value of the EEG signal amplitude after applying the TMS pulse |
| | Skew | Skewness of the EEG signal after applying TMS pulse |
| | Kurtosis | Kurtosis of the EEG signal after applying the TMS pulse |
| Hjorth metrics | Hjorth activity | Variance of the EEG signal after the TMS pulse |
| | Hjorth mobility | Square root of the ratio of the first derivative of the Hjorth activity and the Hjorth activity |
| | Hjorth complexity | Ratio between the Hjorth mobility of the first derivative of the signal and the variance of the signal |
| TEP specific metrics | Energy | Sum of the square root of samples |
| | P1 | Peak value of the EEG between 25 to 40ms after applying the TMS pulse |
| | P2 | Peak value of the EEG between 45 to 80ms after applying the TMS pulse |
| | P3 | Peak value of the EEG between 85 to 150ms after applying the TMS pulse |
| | P4 | Peak value of the EEG between 160 to 250ms after the applying the TMS pulse |
| | AUC GMFP | Area under the curve of the global mean field potential |

Additionally, the area under the curve (AUC) has previously been used to analyze TMS-evoked EEG responses [11]. Here, we calculated the global mean field potential (GMFP) as described in Eq. (1) [12] and extracted the AUC over the entire segment after the pulse.

$$GMFP(t) = \sqrt{\frac{\sum_{j}^{M}(x_j(t) - x_{mean}(t))^2}{M}} \qquad (1)$$

where $t$ - time, $x_j$ – signal from the $j^{th}$ channel, $M$ – number of channels, $x_{mean}$ – mean signal over all channels.

*D. Averaging*

Each feature was extracted for each individual TMS trial. Afterwards, the values were averaged per EEG channel and multiple electrodes. This was not the case for the AUC GMFP, as the metric was computed directly over the entire electrode set. Previous works have analyzed TEP signals as global averages over the entire electrode set or averaged over electrodes in a particular region of interest from the scalp [13]. Here, we performed experiments to test if reduced EEG montages would yield a similar classification performance. Three electrode sets were considered for final feature averaging: (i) high density set – all available electrodes, (ii) medium density set – half of the electrodes available in the 10-10 system, (iii) low density set - based on the locations from the 10-20 electrode positioning system. The channels selected for each in each electrode set are shown in Figure 3.

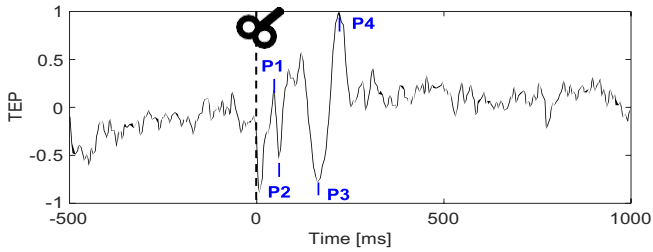

Fig. 2. Example of the four detected TMS-evoked EEG potential peaks (P1: 25-40ms, P2: 45-80ms, P3: 85-150ms, P4: 160-250ms).

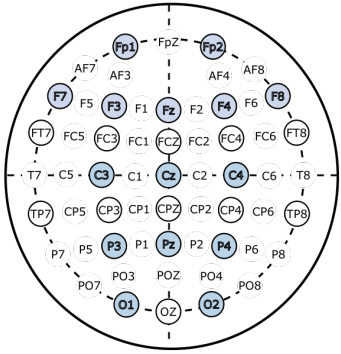

Fig. 3. Electrode channels used in the different averaging experiments. Recordings from all electrodes listed were used in the *high density* set averaging experiments. Electrodes in a black circle were used in the *medium density* set experiment and the electrodes colored in blue were used for the *low density* set averaging experiment.

### E. Classification

The averaged features per patient are used as input to a classification algorithm. We experimented with three types of supervised classifiers:

*Decision Tree (DT)* - a type of rule-based classification that splits the original node based on specific criteria [14]. Several experiments were performed, and entropy was selected as the split quality criteria with the tree having minimum 10 samples per leaf [15]

*k-Nearest Neighbors (kNN)* – classification algorithm that uses the labels of the k nearest training samples to classify the test sample. In our experiments, the best results were obtained with a total of 7 neighbors weighted by the inverse of their distance computed with the ball tree algorithm with a leaf size of 10 based on Euclidian distance [15].

*Random Forests (RF)* – is an ensemble learning method that combines the output of a forest of decision trees through majority voting [16]. A minimum of 10 samples per leaf were selected [15].

TABLE II
DETAILED RESULTS OF THE RF CLASSIFICATION OF AD PATIENTS FROM HC. ACC - ACCURACY, SEN - SENSITIVITY, SPE - SPECIFICITY, $F_1$ - $F_1$ SCORE ALL VALUES ARE EXPRESSED IN PERCENTAGES.

| Experiment | Acc | Sen | Spe | F1 |
|---|---|---|---|---|
| Low density | 69.97 | 65.76 | 74.17 | 69.84 |
| Medium density | 69.47 | 64.70 | 74.23 | 69.32 |
| High density | 92.70 | 96.58 | 88.82 | 92.68 |

### F. Evaluation

As the dataset used for creating the classification model is relatively small, a leave-one-subject-one validation method was selected. In this type of evaluation, the first $k-1$ subjects are used for training and the $k^{th}$ subject is used for testing. The subjects are rotated until all have been tested. The performance of the algorithm is calculated in terms of accuracy, sensitivity, specificity and $F_1$ score:

$$accuracy = (TP + TN)/(FP + FN + TP + TN) \quad (2)$$

$$sensitivty = recall = TP/(TP + FN) \quad (3)$$

$$specificity = TN/(TN + FP) \quad (4)$$

$$precision = TP/(TP + FP) \quad (5)$$

$$F_1 score = 2 precision * recall/(precision + recall) \quad (6)$$

where $TP$ – true positive, $TN$ – true negative, $FP$ – false positive, $FN$ – false negative. To cope with the variability introduced by some of the algorithms, all classification algorithms are run for 100 times. The presented metrics are an average over all runs.

## III. RESULTS

Figure 4 shows the results in terms of accuracy of the classification of Alzheimer's disease patients and healthy controls from all three classifiers. Random Forest shows the highest classification performance with an accuracy of 92.7% for the high-density electrode set. The decision tree model also shows a similarly good performance with 92.2% accuracy. The kNN algorithm has the lowest performance. For all classifiers, the medium and low density electrode sets show a decrease in performance that is 20% lower than the best case scenario.

Table II presents the detailed results of the RF classification, including the sensitivity, specificity and $F_1$ score. The highest values for these metrics are also obtained using features averaged over the high density electrode set with 96.58%, 88.82% and 92.68% for sensitivity, specificity and $F_1$ score respectively. Averaging over the low and medium density electrode sets shows a significant reduction in performance over all metrics.

## IV. DISCUSSION

A high classification performance was obtained for identifying Alzheimer's disease patients from healthy controls using time domain features from TMS-EEG data. The obtained performance is in the same range to values obtained when identifying AD from HC based on other modalities reported in literature. For instance, Bride et. al [17] obtained an accuracy of 95.8% in classifying AD, MCI and HC by applying a support vector machine on Sugihara causality extracted from EEG data. Similarly, Sarica et al [18] obtain a maximum accuracy of 98% using an RF algorithm on neuroimaging data. Compared to other more commonly used modalities for AD diagnosis, TMS-EEG data can additionally provide physicians with functional information as well and could potentially indicate diverse pathological features of altered neuronal communication.

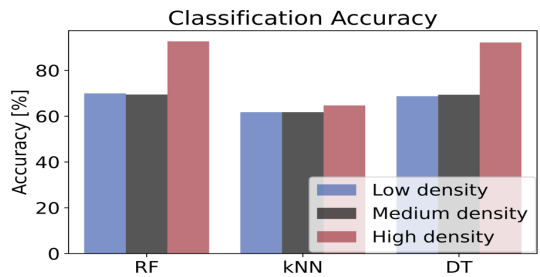

Fig. 4. Accuracy of the classification of AD vs. HC using the three different electrode sets for averaging features and the DT, kNN and RF algorithms for classification.

The RF and DT algorithm performance is similar. The result is expected, as Random Forests are based on decision trees. Both algorithms significantly outperform the kNN classification. This might be due to the distribution of samples in the feature space and the chosen distance metric from the kNN parameter. More experiments would be needed to determine if different distance metrics could improve the performance of kNN.

In all cases, the highest classification performance was obtained when averaging the features over the entire electrode set. Previous literature has shown that globally averaged TMS-EEG responses present a higher differentiation power between AD and HC than locally averaged responses [13]. Therefore, using a higher density electrode montage could be more helpful in detecting the altered brain responses. Nonetheless, the medium and low density montages are spatially distributed over the entire scalp. The lower performance could also be an indication of a poorer signal quality or response to the TMS stimulation. Averaging over a larger set could cancel out bad channel data.

More work is needed to investigate the cause of a better classification performance when using a high-density electrode montage. Frequency content and measures of network connectivity could serve as potential inputs for the classification of AD patients from HC.

## V. CONCLUSION

In this study we proposed a method of classifying Alzheimer disease patients from healthy controls using TMS-EEG responses. Several features were extracted from the time domain signals and averaged over a high, medium and low density electrode set. The highest performance was of 92.7% accuracy obtained using a high-density electrode set for averaging and a random forest classifier.

Further work is needed to evaluate if the results are indicative of altered global brain responses in AD and will involve analysis of EEG frequency and network connectivity. The severity level of the disease should also be considered.